\documentclass[a4paper,fleqn]{cas-dc}

\usepackage[authoryear,longnamesfirst]{natbib}
\usepackage{graphicx}
\usepackage{float}
\usepackage{xcolor}
\def\tsc#1{\csdef{#1}{\textsc{\lowercase{#1}}\xspace}}
\tsc{WGM}
\tsc{QE}
\tsc{EP}
\tsc{PMS}
\tsc{BEC}
\tsc{DE}

\begin{document}
\let\WriteBookmarks\relax
\def\floatpagepagefraction{1}
\def\textpagefraction{.001}

\shorttitle{Teleodynamics and Black Holes}

\shortauthors{Oem Trivedi}

\title [mode = title]{Thermodynamics vs Teleodynamics: A Cosmological Divide?}

\author[1]{Oem Trivedi}[]
\cormark[1]

\ead{oem.trivedi@vanderbilt.edu}

\affiliation[1]{organization={Department of Physics and Astronomy, Vanderbilt University},
    city={Nashville, TN},
    postcode={37235},
    country={USA}}
    
\affiliation[2]{organization={Complex Resilient Intelligent Systems Laboratory, Department of Chemical Engineering, Columbia University}, 
    city={New York, NY},
    postcode={10027}, 
    country={USA}}

\author[2]{Venkat Venkatasubramanian}[]

\ead{venkat@columbia.edu}

\cortext[cor1]{Corresponding author}

\begin{abstract}
We show that black holes and the evolving universe belong to fundamentally different thermodynamic regimes: while stationary black holes obey ordinary Bekenstein–Hawking thermodynamics, cosmology follows memory-bearing teleodynamics, which is also inevitable for non-stationary black holes. This provides a dynamical, semi-classical realization of the thermodynamic split conjecture and identifies memory accumulation as the natural source of deviations from the area law in cosmology. Our results suggest that quantum gravity constructions should not seek to extrapolate black hole thermodynamics to the universe, but instead perhaps incorporate horizon memory as a fundamental microscopic ingredient and consider cosmological constructions consistent with that. 
\end{abstract}


\maketitle

In a recent work, we showed that the teleodynamic extension of statistical mechanics introduced in \citep{tel1Trivedi:2025tmo} can naturally incorporate all the features of modern cosmology. Cosmic Teleodynamics can provide a completely new form of cosmology that explains dark energy, dark matter, and the cosmic tensions by positing a memory-bearing universe. The word "telos" means "goal" in Greek, and is a term which was widely used by Aristotle in his works. This theory generically emerges when one considers cosmology in terms of teleodynamics~\citep{venkat2017book, Venkat2025activematter, venkat2025jaynes, venkat2025social}, introducing game-theoretic concepts of agents to cosmological entities such as galaxies, thereby giving us a memory-bearing regime whose correct coarse-grained description has sharp departures from simple thermodynamic descriptions that have been employed for cosmology so far, by taking the liberty of employing the Bekenstein-Hawking \citep{sb1hawking1972black,sb2bekenstein1973black,sb3bekenstein1974generalized} formalism to the cosmological horizon. This raises a deeper question: are teleodynamics and thermodynamics valid in different regimes? We answer this question in this letter and show that the answer is affirmative. \\

The teleodynamic microscopic ensemble is defined by assigning to each configuration $x$ an energy functional $E(x)$ and a bias functional $\Phi(x)$, with microscopic weight \citep{tel1Trivedi:2025tmo}
\begin{equation}
p(x) = \frac{1}{Z(\beta,\alpha)} \exp\big[-\beta E(x) - \alpha \Phi(x)\big]
\label{eq:px}
\end{equation}
where $\beta$ is the usual inverse temperature parameter, $\alpha$ controls the strength of the teleodynamic bias, and $Z(\beta,\alpha)$ is the generalized partition function~\citep{Venkat2025activematter, venkat2025jaynes}
\begin{equation}
Z(\beta,\alpha) = \sum_x \exp\big[-\beta E(x) - \alpha \Phi(x)\big]
\label{eq:Z}
\end{equation}
The bias $\Phi$ is a scalar functional constructed from coarse-grained geometric and matter invariants. In a covariant formulation, one may write schematically
\begin{equation}
\Phi = \Phi\!\left(\Theta, \sigma_{\mu\nu}\sigma^{\mu\nu}, R, R_{\mu\nu}u^\mu u^\nu, \mathcal{I}_{\rm matter}\right)
\end{equation}
where $\Theta$ is the expansion of a preferred timelike congruence, $\sigma_{\mu\nu}$ its shear, and $\mathcal{I}_{\rm matter}$ denotes coarse-grained matter invariants. Since $\Phi$ is built on scalars, it is invariant under coordinate transformations and slicing choices. In FLRW symmetry, only the spatial average $\bar{\Phi}(t)$ contributes at the background level, whereas fluctuations $\varphi(t,\mathbf{x})$ source inhomogeneous corrections. In this Letter, we keep $\Phi$ general, but restrict attention to scalar covariant constructions, while more explicit realizations are discussed in \citep{tel1Trivedi:2025tmo}.
For histories $\Gamma = \{x(t),p(t)\}$, the maximum-caliber ensemble takes the form \citep{tel1Trivedi:2025tmo}
\begin{equation}
P[\Gamma] \propto \exp\big[-\beta A[\Gamma] - \alpha K[\Gamma]\big]
\label{eq:PGamma}
\end{equation}
where $A[\Gamma]$ is the standard action and $K[\Gamma]$ is a bias functional that encodes memory and environmental dependence, and for collisionless tracers of mass $m$ in an expanding universe with scale factor $a(t)$ and Newtonian potential $\Psi(x,t)$, one may take the following steps.
\begin{equation}
A[\Gamma] = \int dt \left[ \frac{p^2(t)}{2 a^2(t) m} + m\,\Psi(x(t),t) \right]
\label{eq:A}
\end{equation}
and
\begin{equation}
K[\Gamma] = \int \Phi\big(x(t),p(t); f(t,x,p),E(t,x)\big) dt
\label{eq:K}
\end{equation}
with $f(t,x,p)$ being the phase-space distribution and $E(t,x)$ coarse environmental fields. The teleodynamic weight \eqref{eq:PGamma} is not introduced as an arbitrary deformation of the canonical ensemble, but as the natural maximum-caliber extension when additional macroscopic constraints beyond energy conservation are imposed. In a coarse-grained gravitational system with long-range interactions and persistent correlations, the entropy maximization principle must be supplemented by constraints on history-dependent observables. The bias functional \eqref{eq:K} acts as a Lagrange multiplier enforcing such non-equilibrium constraints, and in the limit $\alpha\to 0$ or when $\Phi$ becomes constant over the relevant sector, one recovers the ordinary thermodynamic ensemble. A detailed statistical derivation based on maximum-caliber coarse-graining and large-deviation principles is provided in our companion work on Cosmological Teleodynamics \citep{tel1Trivedi:2025tmo}. 

Ordinary thermodynamics is recovered in a limit whenever the teleodynamic bias is effectively constant over the relevant sector. If $\Phi(x) = \Phi_0$ for all microstates $x$ that contribute to a given macroscopic configuration, then Eq. \eqref{eq:px} becomes
\begin{equation}
p(x) \propto e^{-\beta E(x) - \alpha \Phi_0}
= e^{-\alpha \Phi_0} e^{-\beta E(x)}
\end{equation}
and the constant factor $e^{-\alpha \Phi_0}$ cancels in the normalized ensemble, giving us
\begin{equation}
p(x) = \frac{e^{-\beta E(x)}}{\sum_{x'} e^{-\beta E(x')}}
\label{eq:thermo-limit}
\end{equation}
Similarly, if $K[\Gamma] = T_{\text{obs}}\Phi_0$ is constant along the histories of interest, then we see that Eq. \eqref{eq:PGamma} reduces to
\begin{equation}
P[\Gamma] \propto e^{-\alpha T_{\text{obs}}\Phi_0} e^{-\beta A[\Gamma]}
\end{equation}
and the path ensemble is again purely thermodynamic after normalization.\\

So, what regimes would be where the teleodynamic bias is effectively constant? Stationary black hole spacetimes provide precisely such a regime. Let $(\mathcal{M},g_{\mu\nu})$ be a stationary black hole with a time-like Killing vector $\xi^\mu$ outside the horizon, where the existence of $\xi^\mu$ implies a conserved Hamiltonian $H_K$ and a canonical ensemble
\begin{equation}
Z_{\text{BH}}(\beta) = \text{Tr}\, e^{-\beta H_K}
\end{equation}
with microstate weights $p_i \propto e^{-\beta E_i}$. Here, the teleodynamic bias is a function of conserved charges only, for instance
\begin{equation}
\Phi_i = \Phi(E_i,J_i,Q_i) = \Phi(M,J,Q) \equiv \Phi_0
\end{equation}
for microstates that share the same macroscopic mass $M$, angular momentum $J$, and charge $Q$, Eq. \eqref{eq:px} reduces to Eq. \eqref{eq:thermo-limit}. At the level of histories, orbits of $\xi^\mu$ and their perturbations admit no secular memory accumulation so that the integrand of Eq. \eqref{eq:K} is effectively constant and $K[\Gamma]$ contributes only a normalization. In this stationary Killing horizon regime, teleodynamics reduces to ordinary thermodynamics, and so we see that the Bekenstein–Hawking entropy and its string-theoretic microstate derivations \citep{sv1Strominger:1996sh,sv2Nishioka:2009un,sv3Strominger:1997eq,sv4Bena:2005va} are recovered unchanged. It should be noted that the thermodynamic split here is not merely the statement that time dependence generates corrections, but rather it is the structural claim that a timelike Killing symmetry enforces
\begin{equation}
\mathcal{L}_{\xi}\Phi = 0 \quad \Rightarrow \quad \frac{d}{d\tau}\Phi = 0
\end{equation}
so that the history functional reduces to
\begin{equation}
K[\Gamma]=\Phi_0 T_{\rm obs}
\end{equation}
which factors out of the normalized ensemble, leaving the thermodynamic observables unchanged. The existence of a Killing symmetry therefore guarantees that teleodynamics collapses to ordinary thermodynamics, and in the absence of such a symmetry, $\mathcal{L}_u \Phi\neq 0$ generically, and the accumulated correction
\begin{equation}
\delta K[\Gamma]=\int d\tau \int d\tau' \,\Sigma(\tau')
\end{equation}
remains history-dependent and cannot be absorbed into a normalization. The split is therefore symmetry-protected rather than phenomenologically imposed. \\

Let us make this discussion more explicit and consider that $\xi^\mu$ is a timelike Killing vector that generates stationarity outside the horizon, so that $\mathcal{L}_\xi g_{\mu\nu} = 0$. Along an orbit of $\xi^\mu$, a phase space point $(x^\mu(\tau),p_\mu(\tau))$ evolves according to
\begin{equation}
\frac{d x^\mu}{d\tau} = \xi^\mu(x(\tau)), 
\qquad 
\frac{d p_\mu}{d\tau} = \mathcal{F}_\mu(x(\tau),p(\tau))
\end{equation}
where $\tau$ is a parameter adapted to the Killing flow and $\mathcal{F}_\mu$ encodes the usual Hamiltonian evolution, and so here in the stationary black hole regime, the relevant coarse-grained fields $f$ and $E$ entering the bias functional satisfy
\begin{equation}
\mathcal{L}_\xi f = 0, 
\qquad 
\mathcal{L}_\xi E = 0
\end{equation}
That is, they are invariant under the Killing flow. Note that the bias density $\Phi(x,p;f,E)$ is constructed from $f$, $E$, and local geometric invariants, so Killing invariance implies
\begin{equation}
\mathcal{L}_\xi \Phi = \xi^\mu \nabla_\mu \Phi = 0
\end{equation}
This, in turn, ends up giving us
\begin{equation}
\frac{d}{d\tau} \Phi\big(x(\tau),p(\tau);f,E\big) 
= \xi^\mu \nabla_\mu \Phi = 0
\end{equation}
Therefore, along any orbit of $\xi^\mu$ one has
\begin{equation}
\Phi\big(x(\tau),p(\tau);f,E\big) = \Phi_0
\end{equation}
This is a constant that depends only on the black hole's conserved macroscopic charges, not on its microscopic trajectory details. The bias functional is then given as
\begin{equation}
K[\Gamma] 
= \int_{\tau_i}^{\tau_f} d\tau\; \Phi\big(x(\tau),p(\tau);f,E\big)
= \Phi_0\,(\tau_f - \tau_i)
\end{equation}
For an observation time $T_{\text{obs}} = \tau_f - \tau_i$, this gives us the following
\begin{equation}
K[\Gamma] = \Phi_0\, T_{\text{obs}}
\end{equation}
which is independent of the detailed history $\Gamma$ as long as one remains in the stationary sector. The teleodynamic weight for histories then becomes
\begin{equation}
P[\Gamma] \propto \exp\big[-\beta A[\Gamma] - \alpha \Phi_0 T_{\text{obs}}\big]
= e^{-\alpha \Phi_0 T_{\text{obs}}}\, e^{-\beta A[\Gamma]}
\end{equation}
so that all histories in the stationary sector acquire the same multiplicative factor $e^{-\alpha \Phi_0 T_{\text{obs}}}$. After normalization, this factor will cancel out, and the normalized path probability reduces to
\begin{equation}
P[\Gamma] = \frac{e^{-\beta A[\Gamma]}}{\int \mathcal{D}\Gamma'\, e^{-\beta A[\Gamma']}}
\end{equation}
which is exactly the standard thermodynamic path ensemble. We can also reach the same conclusion at the level of the partition function, where the teleodynamic partition function factorizes as
\begin{multline}
Z(\beta,\alpha) 
= \sum_{\Gamma} \exp\big[-\beta A[\Gamma] - \alpha K[\Gamma]\big]
= \\ e^{-\alpha \Phi_0 T_{\text{obs}}} \sum_{\Gamma} e^{-\beta A[\Gamma]}
= e^{-\alpha \Phi_0 T_{\text{obs}}} Z_{\text{BH}}(\beta)    
\end{multline}
where $Z_{\text{BH}}(\beta)$ is the usual black hole partition function, and the free energy and entropy also follow as
\begin{equation}
F(\beta,\alpha) = -\frac{1}{\beta} \ln Z(\beta,\alpha)
= -\frac{1}{\beta} \ln Z_{\text{BH}}(\beta) + \frac{\alpha \Phi_0 T_{\text{obs}}}{\beta},
\end{equation}
\begin{equation}
S(\beta,\alpha) = \left(1 - \beta \frac{\partial}{\partial \beta}\right) \ln Z(\beta,\alpha)
= \left(1 - \beta \frac{\partial}{\partial \beta}\right) \ln Z_{\text{BH}}(\beta)
\end{equation}
since the additive term $\alpha \Phi_0 T_{\text{obs}}$ is independent of $\beta$. Thus, what we see is that the macroscopic entropy and all related thermodynamic quantities coincide with those obtained from $Z_{\text{BH}}(\beta)$ alone. In particular, the Bekenstein–Hawking entropy and its string theoretic microstate derivations are unchanged by the teleodynamic factor in the stationary Killing-horizon regime, confirming that teleodynamics reduces to ordinary thermodynamics in this limit.\\

Cosmology, by contrast, is intrinsically time-dependent and non-equilibrium, as for a spatially flat FLRW universe with scale factor $a(t)$ and Hubble rate $H(t) = \dot a/a$, there is no global timelike Killing vector and the large-scale matter distribution develops a nontrivial cosmic web of filaments, walls, and voids. In this setting, the teleodynamic bias must be decomposed as
\begin{equation}
\Phi(t,x) = \bar{\Phi}(t) + \varphi(t,x)
\label{eq:phi-split}
\end{equation}
with $\bar{\Phi}(t)$ the spatial average and $\varphi(t,x)$ the fluctuation. As established in \citep{tel1Trivedi:2025tmo}, the homogeneous part gives rise to an effective teleodynamic energy density
\begin{equation}
\rho_{\text{TD}}(t) = \alpha\,\bar{\Phi}(t)
\label{eq:rhoTD}
\end{equation}
leading to a modified Friedmann equation,
\begin{equation}
3 M_P^2 H^2(t) = \rho(t) + \rho_{\text{TD}}(t)
\label{eq:Friedmann-TD}
\end{equation}
where $\rho(t)$ denotes the usual matter and radiation energy density. The associated pressure $p_{\text{TD}}(t)$ is obtained from the time dependence of $\bar{\Phi}(t)$ and produces an effective equation of state $w_{\text{TD}}(t) = p_{\text{TD}}(t)/\rho_{\text{TD}}(t)$ that can mimic a cosmological constant-like, quintessence-like, or phantom-like behavior depending on the evolution of $\bar{\Phi}(t)$. Note also that in a covariant formulation, the teleodynamic contribution may be incorporated through an effective action \citep{tel1Trivedi:2025tmo}
\begin{equation}
S_{\rm eff}=\int d^4x \sqrt{-g}\left(\frac{M_P^2}{2}R+\mathcal{L}_m\right)
-\alpha \int d^4x \sqrt{-g}\,\Phi
\end{equation}
which gives
\begin{equation}
M_P^2 G_{\mu\nu}=T^{(m)}_{\mu\nu}+T^{({\rm TD})}_{\mu\nu},
\end{equation}
with
\begin{equation}
T^{({\rm TD})}_{\mu\nu}=\alpha \Phi g_{\mu\nu}-2\alpha \frac{\delta\Phi}{\delta g^{\mu\nu}}.
\end{equation}
The Bianchi identities ensure $\nabla_\mu T^{\mu\nu}_{\rm eff}=0$ once the metric dependence of $\Phi$ is specified, so the modified Friedmann system is self-consistent. 

The inhomogeneous part $\varphi(t,x)$ modifies the Boltzmann and Poisson equations. The teleodynamic Boltzmann equation for the distribution function $f(t,x,p)$ in comoving coordinates reads \citep{tel1Trivedi:2025tmo}
\begin{equation}
\frac{\partial f}{\partial t} + \frac{p}{a^2 m} \cdot \nabla_x f
- \left( \nabla_x \Psi + \alpha \nabla_x \Phi \right) \cdot \nabla_p f
- H\, p \cdot \nabla_p f = 0
\label{eq:boltzmann-TD}
\end{equation}
with $\Psi(t,x)$ the Newtonian potential, and the teleodynamic term introduces an additional drift $-\alpha\nabla_x\Phi$, corresponding to an effective potential
\begin{equation}
\Psi_{\text{eff}}(t,x) = \Psi(t,x) + \alpha\,\varphi(t,x)
\label{eq:Psi-eff}
\end{equation}
since $\bar{\Phi}(t)$ does not contribute to spatial gradients, and the Poisson equation becomes
\begin{equation}
\nabla^2 \Psi_{\text{eff}}(t,x) = 4\pi G a^2(t)\, \left[ \bar{\rho}(t)\,\delta(t,x) + \rho_{\text{TD}}(t,x) \right]
\label{eq:poisson-TD}
\end{equation}
with
\begin{equation}
\nabla^2 \Psi(t,x) = 4\pi G a^2(t)\, \bar{\rho}(t)\,\delta(t,x)
\end{equation}
and
\begin{equation}
\rho_{\text{TD}}(t,x) = \frac{\alpha}{4\pi G a^2(t)}\, \nabla^2 \varphi(t,x)
\label{eq:rhoTD-local}
\end{equation}
In Fourier space, one can posit a teleodynamic response kernel $K(k,a)$ such that
\begin{equation}
\varphi_k(a) = K(k,a)\, \delta_k(a)
\end{equation}
which gives us the emergent clustering density,
\begin{equation}
\rho_{\text{TD}}(k,a) = \Delta\mu(k,a)\, \bar{\rho}(a)\, \delta_k(a)
\end{equation}
with
\begin{equation}
\Delta\mu(k,a) = \frac{\alpha K(k,a)}{4\pi G a^2}
\label{eq:deltamu}
\end{equation}
The linear growth equation for the density contrast then takes the form
\begin{equation}
\ddot{\delta}_k + 2 H \dot{\delta}_k
- 4\pi G \bar{\rho}\,[1 + \Delta\mu(k,a)]\,\delta_k = 0
\label{eq:growth-TD}
\end{equation}
Equations \eqref{eq:Friedmann-TD} and \eqref{eq:growth-TD} express the central teleodynamic fact that the same functional $\Phi(t,x)$, through its homogeneous and inhomogeneous parts, drives both an effective dark energy sector and a scale-dependent modification of clustering. In a universe with evolving large-scale structure, correlators and tidal fields, $\bar{\Phi}(t)$ and $\varphi(t,x)$ cannot be treated as constants. The teleodynamic corrections are strong here, in sharp contrast to the stationary black hole case, where $\Phi$ may consistently be taken as constant over the relevant ensemble. Therefore, it should be noted that equations (\ref{eq:boltzmann-TD})-(\ref{eq:growth-TD}) form a closed linear system once a response kernel $K(k,a)$ is specified. The teleodynamic sector therefore modifies growth only through the scale-dependent factor $\Delta\mu(k,a)$, while preserving the standard structure of the Boltzmann hierarchy and metric perturbation equations. In the limit $\alpha\to 0$ (or $K\to 0$), one recovers the standard $\Lambda$CDM growth equation, demonstrating that teleodynamics acts as a controlled deformation of linear perturbation theory rather than a structural replacement. \\

But now consider a different type of black hole scenario. Let us consider a black hole embedded in a time-dependent, cosmologically expanding universe. In this situation, there is no timelike Killing vector field that generates an exact stationarity of the geometry in the near-horizon region.  So, any attempt to impose an exactly static event horizon
in a time-dependent background turns that surface into a naked null singularity\citep{non1Farrah:2023opk,non2Faraoni:2024ghi}. Instead, we work with a physically
preferred timelike vector field $u^\mu$ describing the congruence of observers adapted to the cosmologically coupled horizon, where, for instance, the observers
following the quasi-local horizon or asymptotically comoving with the FLRW background. Along an integral curve of $u^\mu$, a phase space point
$(x^\mu(\tau),p_\mu(\tau))$ evolves according to
\begin{equation}
\frac{d x^\mu}{d\tau} = u^\mu(x(\tau)),
\qquad 
\frac{d p_\mu}{d\tau} = \mathcal{F}_\mu\big(x(\tau),p(\tau);\tau\big),
\label{eq:cosmo-bh-flow}
\end{equation}
where $\tau$ is the proper time parameter along $u^\mu$ and
$\mathcal{F}_\mu$ encodes the usual Hamiltonian evolution together with any slow time dependence induced by the cosmological expansion in the near-horizon geometry. Now, in the cosmologically coupled regime, the metric, the coarse-grained distribution $f$, and the environmental fields $E$ entering the bias functional are no longer invariant along $u^\mu$. So instead of the Killing
invariance we have
\begin{equation}
\mathcal{L}_u g_{\mu\nu} \neq 0, 
\qquad 
\mathcal{L}_u f \neq 0, 
\qquad 
\mathcal{L}_u E \neq 0
\label{eq:cosmo-bh-nonstationary}
\end{equation}
This goes towards reflecting the fact that the horizon radius, the ambient curvature, and the cosmic environment evolve with cosmological time even in the vicinity of the
black hole. \\ 

The teleodynamic bias density $\Phi(x,p;f,E)$ is constructed from $f$, $E$, and local geometric invariants (including the evolving near-horizon curvature), where note that in the stationary case, Killing invariance implies
$\mathcal{L}_\xi \Phi = 0$ and a constant bias along $\xi^\mu$. But in the cosmologically coupled case, here, the lack of such a symmetry generically implies
\begin{equation}
\mathcal{L}_u \Phi 
= u^\mu \nabla_\mu \Phi(x,p;f,E)
\equiv \Sigma(x,p;f,E) \neq 0
\label{eq:cosmo-bh-sigma-def}
\end{equation}
where we have defined $\Sigma$ as the teleodynamic ``source" encoding the
build up of memory and environmental dependence along the non-stationary
flow. Equation \eqref{eq:cosmo-bh-sigma-def} translates along an orbit of
$u^\mu$ into
\begin{equation}
\frac{d}{d\tau} \Phi\big(x(\tau),p(\tau);f,E\big) 
= u^\mu \nabla_\mu \Phi
= \Sigma\big(x(\tau),p(\tau);f,E\big)
\label{eq:cosmo-bh-dPhi-dtau}
\end{equation}
which is generically non-zero in a time-dependent near-horizon environment. We can then continue by integrating Eq.~\eqref{eq:cosmo-bh-dPhi-dtau} along the history from
$\tau_i$ to $\tau_f$ to get
\begin{multline}
\Phi\big(x(\tau),p(\tau);f,E\big)
= \Phi_0 + \delta\Phi(\tau) \\
\delta\Phi(\tau)
= \int_{\tau_0}^{\tau} d\tau'\,
\Sigma\big(x(\tau'),p(\tau');f,E\big)
\label{eq:cosmo-bh-Phi-decomp}    
\end{multline}

where $\Phi_0 \equiv \Phi(\tau_0)$ is a reference value, which may depend on the macroscopic charges $(M,J,Q)$ and possibly on the asymptotic cosmological parameters, and $\delta\Phi(\tau)$ captures the accumulated teleodynamic
memory along the non-stationary flow. The bias functional $K[\Gamma]$ is then
\begin{align}
K[\Gamma] 
&= \int_{\tau_i}^{\tau_f} d\tau\; 
\Phi\big(x(\tau),p(\tau);f,E\big)
\nonumber\\[4pt]
&= \int_{\tau_i}^{\tau_f} d\tau\; 
\Big[\Phi_0 + \delta\Phi(\tau)\Big]
\nonumber\\[4pt]
&= \Phi_0\,(\tau_f - \tau_i)
+ \int_{\tau_i}^{\tau_f} d\tau\; \delta\Phi(\tau)
\nonumber\\[4pt]
&= \Phi_0\,T_{\text{obs}} + \delta K[\Gamma]
\label{eq:cosmo-bh-K-split}
\end{align}
where $T_{\text{obs}} \equiv \tau_f - \tau_i$ is the observation time. We have defined the genuinely teleodynamic history-dependent correction as
\begin{equation}
\delta K[\Gamma]
\equiv \int_{\tau_i}^{\tau_f} d\tau\; \delta\Phi(\tau)
= \int_{\tau_i}^{\tau_f} d\tau \int_{\tau_0}^{\tau} d\tau'\,
\Sigma\big(x(\tau'),p(\tau');f,E\big)
\label{eq:cosmo-bh-deltaK-def}
\end{equation}
In this case, which is in stark contrast to the stationary Killing horizon regime, $\delta K[\Gamma]$ does
not reduce to a constant but depends on the detailed history $\Gamma$, through
the time-dependent accumulation of teleodynamic bias sourced by
$\Sigma\neq 0$. \\

The teleodynamic weight for histories in the cosmologically coupled regime thus becomes
\begin{align}
P[\Gamma] 
&\propto \exp\big[-\beta A[\Gamma] - \alpha K[\Gamma]\big]
\nonumber\\[4pt]
&= \exp\big[-\beta A[\Gamma] - \alpha \Phi_0 T_{\text{obs}}
           - \alpha \delta K[\Gamma]\big]
\nonumber\\[4pt]
&= e^{-\alpha \Phi_0 T_{\text{obs}}}\,
   \exp\big[-\beta A[\Gamma] - \alpha \delta K[\Gamma]\big]
\label{eq:cosmo-bh-path-weight}
\end{align}
While the multiplicative factor $e^{-\alpha \Phi_0 T_{\text{obs}}}$ is the
same for all histories in the sector with fixed macroscopic charges, the second factor now contains the non-trivial, history-dependent functional
$\delta K[\Gamma]$. After normalization, we see that the constant factor cancels as
before, but the $\delta K[\Gamma]$ dependence survives, which gives us
\begin{equation}
P[\Gamma] 
= \frac{
\exp\big[-\beta A[\Gamma] - \alpha \delta K[\Gamma]\big]
}{
\displaystyle\int \mathcal{D}\Gamma'\,
\exp\big[-\beta A[\Gamma'] - \alpha \delta K[\Gamma']\big]
}
\label{eq:cosmo-bh-normalized-path}
\end{equation}
This is a striking result, as this is no longer the standard thermodynamic path ensemble, which means that the teleodynamic bias does not collapse to a trivial overall normalization, but remains a
genuinely dynamical ingredient of the coarse-grained description. \\

At the level of the partition function, the teleodynamic partition function for a cosmologically coupled black hole reads as
\begin{align}
Z(\beta,\alpha) 
&= \sum_{\Gamma} \exp\big[-\beta A[\Gamma] - \alpha K[\Gamma]\big]
\nonumber\\[4pt]
&= \sum_{\Gamma} 
\exp\big[-\beta A[\Gamma] - \alpha \Phi_0 T_{\text{obs}}
          - \alpha \delta K[\Gamma]\big]
\nonumber\\[4pt]
&= e^{-\alpha \Phi_0 T_{\text{obs}}}
   \sum_{\Gamma} \exp\big[-\beta A[\Gamma] - \alpha \delta K[\Gamma]\big]
\nonumber\\[4pt]
&\equiv e^{-\alpha \Phi_0 T_{\text{obs}}}\, 
        \widetilde{Z}(\beta,\alpha)
\label{eq:cosmo-bh-Ztilde}
\end{align}
where we have defined the reduced partition function as
$\widetilde{Z}(\beta,\alpha)$, with this function incorporating the nontrivial teleodynamic
corrections. In general, we can say that $\widetilde{Z}(\beta,\alpha)$ does not factorize into a purely thermodynamic part and a simple $\alpha$-dependent
normalization, because $\delta K[\Gamma]$ depends on $\Gamma$. The free energy and entropy are then given by
\begin{equation}
F(\beta,\alpha) 
= -\frac{1}{\beta} \ln Z(\beta,\alpha)
= -\frac{1}{\beta} \ln \widetilde{Z}(\beta,\alpha) 
  + \frac{\alpha \Phi_0 T_{\text{obs}}}{\beta}
\label{eq:cosmo-bh-free-energy}
\end{equation}
\begin{equation}
S(\beta,\alpha) 
= \left(1 - \beta \frac{\partial}{\partial \beta}\right) 
  \ln Z(\beta,\alpha)
= \left(1 - \beta \frac{\partial}{\partial \beta}\right) 
  \ln \widetilde{Z}(\beta,\alpha)
\label{eq:cosmo-bh-entropy}
\end{equation}
Note that here, unlike in the stationary Killing horizon case, $\ln \widetilde{Z}(\beta,\alpha)$ carries non-trivial $\alpha$ dependent corrections through the history dependent $\delta K[\Gamma]$. As a result, the macroscopic entropy and related thermodynamic quantities receive genuine
teleodynamic contributions that a simple redefinition of the partition function cannot remove. \\

In the formal limit in which the cosmological coupling is turned off, we see that the near-horizon geometry admits an exact timelike Killing vector field, wherein the
source $\Sigma$ in Eq.~\eqref{eq:cosmo-bh-sigma-def} vanishes
$\delta\Phi(\tau)\to 0$ and $\delta K[\Gamma]\to 0$, so that
$\widetilde{Z}(\beta,\alpha)\to Z_{\text{BH}}(\beta)$ and one recovers the
stationary result. Cosmologically coupled black holes, which one may think of as "realistic" black holes, given that we live in an expanding universe, therefore interpolate
between the pure Bekenstein-Hawking thermodynamic regime and a genuinely
teleodynamic regime in which horizon memory and non-equilibrium effects
become strong. \\ 

But now let us go to the deeper subtleties of this calculation. From a physical and game-theoretic perspective, the appearance of teleodynamic corrections in cosmologically coupled black holes signifies that such objects are no longer isolated, memoryless equilibrium systems, but active participants in the evolving cosmic environment. In the language of teleodynamics, a black hole embedded in an expanding universe experiences a non-constant bias functional due to the time-dependent large-scale fields, tidal structures, and horizon geometry generated by cosmic evolution. This means that realistic black holes accumulate a form of "environmental memory", which is, in a sense, analogous to other teleodynamic agents, as their coarse-grained behavior is influenced not only by local, conserved charges $(M,J,Q)$ but also by the global expansion history and the evolving matter distribution around them. \\

In this sense, a cosmologically coupled black hole responds teleodynamically to the expansion of the universe, developing a non-trivial bias functional that encodes how its past interactions and the surrounding cosmic background shape its effective thermodynamics. Such black holes thus possess a teleodynamic "memory bias" that modifies their horizon statistics, indicating that they lie in the same non-equilibrium, memory-bearing regime as the universe itself rather than in the strictly stationary sector where Bekenstein-Hawking thermodynamics applies exactly. So, one needs to consider a game-theoretic version of thermodynamics even for such realistic black holes. \\ 

Horizon thermodynamics can make this distinction even more transparent. For that, consider the apparent horizon of radius $r_H(t) = H^{-1}(t)$ in a spatially flat FLRW universe. Following \citep{tel1Trivedi:2025tmo}, we model the horizon as composed of $N(H)$ coarse-grained cells of correlation length $\ell_\chi$, so that
\begin{equation}
N(H) = \frac{A}{\ell_\chi^2} = \frac{4\pi}{\ell_\chi^2 H^2}
\end{equation}
where $A = 4\pi r_H^2 = 4\pi H^{-2}$ is the horizon area, and if each cell carries an entropy $s_{\text{cell}}$, the geometric teleodynamic entropy is
\begin{equation}
S_{\text{geom}}(H) = N(H)\, s_{\text{cell}} = \frac{4\pi s_{\text{cell}}}{\ell_\chi^2}\, \frac{1}{H^2} \equiv \frac{C_{\text{TD}}}{H^2}
\label{eq:Sgeom}
\end{equation}
with
\begin{equation}
C_{\text{TD}} = \frac{4\pi s_{\text{cell}}}{\ell_\chi^2}
\end{equation}
In a non-equilibrium teleodynamic setting, there is, in addition, a history-dependent entropy production term, so that the total teleodynamic horizon entropy is
\begin{equation}
S_{\text{TD}}(t) = \frac{C_{\text{TD}}(t)}{H^2(t)} + \int^t \sigma_{\text{TD}}(t')\, dt'
\label{eq:STD}
\end{equation}
where $\sigma_{\text{TD}}(t) \geq 0$ is the entropy production rate associated with teleodynamic memory and fluxes. It should now be noted that quantity $C_{\rm TD}(t)$ encodes the coarse-grained microstate density per correlation area and is defined geometrically once the correlation scale $\ell_\chi$ and cell entropy $s_{\rm cell}$ are specified. The entropy production rate $\sigma_{\rm TD}(t)$ represents non-equilibrium memory accumulation and is defined through the Clausius balance relation. Thus, $C_{\rm TD}$ characterizes the equilibrium microstructure, while $\sigma_{\rm TD}$ parameterizes departures from equilibrium; both reduce to constants (with $\sigma_{\rm TD}=0$) in the stationary/Killing-horizon limit. A detailed derivation of the modified horizon entropy balance used here is given in our companion paper \citep{tel1Trivedi:2025tmo}, where the construction is developed explicitly from a Clausius-type relation on the apparent horizon with a well-defined heat flux written in terms of the quasi-local/Misner-Sharp energy inside the Hubble sphere, together with a non-equilibrium entropy production term that encodes teleodynamic memory. The Misner–Sharp energy inside the Hubble sphere is
\begin{equation}
E_H(t) = \rho_{\text{eff}}(t)\, V_H(t),
\end{equation}
with $V_H(t) = \frac{4\pi}{3} r_H^3 = \frac{4\pi}{3} H^{-3}$ and $\rho_{\text{eff}}(t) = 3 M_P^2 H^2(t)$ and this gives
\begin{equation}
E_H(t) = \frac{4\pi}{3} H^{-3}(t)\, (3 M_P^2 H^2(t)) = 4\pi M_P^2\, \frac{1}{H(t)},
\label{eq:EH}
\end{equation}
and hence
\begin{equation}
\frac{dE_H}{dH} = -\frac{4\pi M_P^2}{H^2}
\label{eq:dEHdH}
\end{equation}
Differentiating Eq. \eqref{eq:STD} with respect to time yields
\begin{equation}
\dot S_{\text{TD}} = -\frac{2 C_{\text{TD}} \dot H}{H^3} + \frac{\dot C_{\text{TD}}}{H^2} + \sigma_{\text{TD}}(t)
\label{eq:dSTDdt}
\end{equation}
Note here that the entropy production rate $\sigma_{\rm TD}\ge 0$ is fixed by construction as the non-equilibrium contribution associated with memory accumulation in the maximum-caliber ensemble. In expanding FLRW backgrounds with $\dot H<0$ and slowly varying $C_{\rm TD}$, the geometric term is positive and together with $\sigma_{\rm TD}\ge 0$ ensures $\dot S_{\rm TD}\ge 0$. The stationary/Killing limit corresponds to $\sigma_{\rm TD}=0$ and the constant $C_{\rm TD}$. The teleodynamic temperature is defined as
\begin{equation}
T_{\text{TD}}^{-1} = \frac{\partial S_{\text{TD}}}{\partial E_H}
= \frac{dS_{\text{TD}}/dH}{dE_H/dH}
\label{eq:TTD-def}
\end{equation}
For slowly varying $C_{\text{TD}}$ and neglecting the explicit $H$ dependence of the integral term in Eq. \eqref{eq:STD}, one finds
\begin{equation}
\frac{dS_{\text{TD}}}{dH} \simeq -\frac{2 C_{\text{TD}}}{H^3}
\end{equation}
and combining this with Equation \eqref{eq:dEHdH} gives
\begin{equation}
\frac{\partial S_{\text{TD}}}{\partial E_H} \simeq \frac{C_{\text{TD}}}{2\pi M_P^2} \frac{1}{H}
\end{equation}
This gives us
\begin{equation}
T_{\text{TD}}(H) \simeq \frac{2\pi M_P^2}{C_{\text{TD}}}\, H
\label{eq:TTD}
\end{equation}
For the special calibration $C_{\text{TD}} = 8\pi^2 M_P^2$ Eq. \eqref{eq:TTD} reduces to $T_{\text{TD}} = H/(2\pi)$, reproducing the Gibbons–Hawking temperature. \\

To obtain the dynamical content, we impose a Clausius-type relation on the apparent horizon,
\begin{equation}
\dot Q = T_{\text{TD}}\, \dot S_{\text{TD}}
\label{eq:Clausius}
\end{equation}
with a flux
\begin{equation}
\dot Q = -\frac{4\pi}{H^2} (\rho_{\text{eff}} + p_{\text{eff}}),
\label{eq:Qdot}
\end{equation}
where $\rho_{\text{eff}}$ and $p_{\text{eff}}$ include both conventional and teleodynamic contributions, and then, substituting Eqs. \eqref{eq:TTD} and \eqref{eq:dSTDdt} into Eq. \eqref{eq:Clausius} and using Eq. \eqref{eq:Qdot}, one finds
\begin{equation}
- \frac{4\pi}{H^2} (\rho_{\text{eff}} + p_{\text{eff}}) = \frac{2\pi M_P^2}{C_{\text{TD}}} H \left( -\frac{2 C_{\text{TD}} \dot H}{H^3} + \frac{\dot C_{\text{TD}}}{H^2} + \sigma_{\text{TD}} \right)
\end{equation}
This simplifies to
\begin{equation}
-(\rho_{\text{eff}} + p_{\text{eff}}) = M_P^2 \left( -2 \dot H + H \frac{\dot C_{\text{TD}}}{C_{\text{TD}}} + H^3 \frac{\sigma_{\text{TD}}}{C_{\text{TD}}} \right)
\end{equation}
This can equivalently be
\begin{equation}
\dot H = -\frac{\rho_{\text{eff}} + p_{\text{eff}}}{2 M_P^2} + \frac{H}{2} \frac{\dot C_{\text{TD}}}{C_{\text{TD}}} + \frac{H^3}{2 C_{\text{TD}}} \sigma_{\text{TD}}
\label{eq:Raychaudhuri-TD}
\end{equation}
Equation \eqref{eq:Raychaudhuri-TD} is the teleodynamic Raychaudhuri equation. In the stationary black hole limit, one expects no net entropy production and a fixed microstate density per correlation area, so that
\begin{equation}
\dot C_{\text{TD}} = 0, \qquad \sigma_{\text{TD}} = 0
\end{equation}
and Eq. \eqref{eq:Raychaudhuri-TD} reduces to
\begin{equation}
\dot H = -\frac{\rho_{\text{eff}} + p_{\text{eff}}}{2 M_P^2}
\end{equation}
This is the usual GR identity that underlies the thermodynamic derivations of Einstein’s equations. With the calibration $C_{\text{TD}} = 8\pi^2 M_P^2$, the familiar Gibbons–Hawking entropy and temperature are recovered \citep{gh1Gibbons:1977mu,gh2Gibbons:1976ue,gh3Gibbons:1978ac,gh4Hawking:1995fd,ha1Hayward:1997jp,ha2Hayward:2008jq}. To be even more clear, in the equilibrium (stationary/Killing-horizon) limit, the microstate density per correlation area is time-independent and there is no net entropy production so $\dot C_{\rm TD}\rightarrow 0$ and $\sigma_{\rm TD}\rightarrow 0$, and the horizon balance collapses to the usual GR identity underlying standard horizon thermodynamics. In exact de Sitter, $H=\mathrm{const.}$ and the same equilibrium conditions render $S_{\rm TD}$ a state function and in quasi dS/slow-roll the teleodynamic corrections are parametrically small, controlled by $\dot C_{\rm TD}/(H C_{\rm TD})$ and $\sigma_{\rm TD}/(H^{-3}C_{\rm TD})$, providing a controlled expansion around the standard Gibbons-Hawking sector. Stationary black hole thermodynamics, therefore, appears as the equilibrium, constant $C_{\text{TD}}$, zero $\sigma_{\text{TD}}$ limit of teleodynamic horizon thermodynamics. \\

Cosmology generically lies away from this equilibrium limit as the Hubble rate evolves, correlation lengths, and cell entropies change as large-scale structure forms, and teleodynamic memory accumulates. In this regime, $\dot C_{\text{TD}}$ and $\sigma_{\text{TD}}$ cannot both vanish without eliminating the very physics responsible for dark energy, clustering modifications, and horizon memory. The teleodynamic corrections in Eq. \eqref{eq:Raychaudhuri-TD} thus become prevalent, but necessary consequences of the non-equilibrium, history-dependent character of the universe. \\

The combination of Eqs. \eqref{eq:Friedmann-TD}, \eqref{eq:growth-TD}, and \eqref{eq:Raychaudhuri-TD} establishes a sharp distinction, which is that black holes live in the equilibrium teleodynamic regime where the bias functional is constant and Bekenstein–Hawking thermodynamics is valid, while cosmology inhabits a non-equilibrium regime where the correct coarse-grained theory is teleodynamic. Deviations of cosmological horizon entropy from the black hole area law are then naturally interpreted as teleodynamic memory corrections governed by $C_{\text{TD}}(t)$ and $\sigma_{\text{TD}}(t)$, while other formulations like Tsallis or Barrow entropies \citep{Tsallis:2012js,entBarrow:2020tzx} are rooted non-additive extensions and quantum gravitational phenomena. It should also be noted that there are interesting recent works \citep{cit1Tsilioukas:2023tdw,cit2Tsilioukas:2024seh} which have considered the Universe horizon being affected by topology changes, a phenomenon that in principle could also be incorporated in the present formalism. If future observations detect departures from Gibbons–Hawking scaling that track the growth of large-scale structure and horizon memory, the simplest explanation will not be a static modification of the area law, but the fact that the universe obeys teleodynamics while black holes obey thermodynamics. \\

We also note that non-equilibrium horizon thermodynamics has previously been formulated through entropy production terms in modified gravity and Jacobson-type derivations \citep{cht1Jacobson:1995ab,jac2Eling:2006aw,jac3Wald:1993nt,cai2008corrected}. The present framework differs in that the entropy production term arises from a microscopic statistical bias in the path ensemble rather than from a modification of the gravitational Lagrangian. In this sense, teleodynamics is not a repackaging of bulk-viscosity corrections, but a statistical generalization of the ensemble itself as the modified Friedmann and perturbation equations emerge from the same bias functional that governs the horizon entropy, providing a unified origin for background acceleration and clustering corrections. \\

From a quantum gravity perspective, the distinction between black hole thermodynamics and cosmological teleodynamics carries structural significance, as Black hole thermodynamics arises in settings where a Killing horizon provides a stationary environment, allowing a microcanonical sector with well-defined charges $(M,J,Q)$ and a fixed surface gravity $\kappa$. In such a regime, the teleodynamic bias functional reduces to a constant, as shown by $\mathcal{L}_\xi \Phi = 0$, so that $K[\Gamma] = \Phi_0 T_{\rm obs}$ contributes only a normalization to the path ensemble. This collapse of teleodynamics into ordinary thermodynamics yields the Bekenstein–Hawking law $S_{\rm BH} = A/4G$ and also underlies the success of string-theoretic microstate counting, the quantum entropy function, and related constructions. These frameworks succeed precisely because the microscopic Hilbert space is organized into nearly time-independent sectors whose degeneracies are protected by supersymmetry, modular invariance, and conserved charges \citep{sv5Carlip:1998wz,sv6DeHaro:2019gno,sv7Emparan:2006it}. In this sense, the thermodynamic split conjecture reflects a QG-relevant structural divide - that stationary horizons admit a temperature $T=\kappa/2\pi$ fixed by Killing symmetry - while cosmological horizons lack the stationary environment required for such a microstate interpretation. \\

Cosmology, on the other hand, is inherently non-stationary and memory-bearing in our framework. This semi-classical teleodynamic memory is the simplest possible source of deviations from the standard thermodynamic area law. The horizon entropy $S_{\rm hor}(t) \propto H^{-2}(t)$ already evolves because $\dot H\neq 0$, but the full teleodynamic entropy includes additional non-equilibrium contributions, $S_{\rm TD}(t)=C_{\rm TD}(t)/H^2(t)+\int^t dt'\,\sigma_{\rm TD}(t')$, which any stationary microstate counting cannot reproduce. If the universe exhibits non-vanishing entropy production $\sigma_{\rm TD}>0$ or a time-varying microstate density $C_{\rm TD}(t)$ driven by large-scale structure and horizon memory, then any attempt to impose a black hole-like quantum gravity formula $S=A/4G$ on cosmology should consider these issues. The quantum gravity implication here is clear: instead of extending black hole thermodynamics to cosmology, a successful quantum theory of cosmological horizons might also need to incorporate memory, coarse-graining, and non-equilibrium structure at a fundamental level. The teleodynamic framework shows that the smallest, most "classical" departure from ordinary thermodynamics already generates new cosmological horizon laws. Thus, if one is to consider the notions of both cosmic teleodynamics and thermodynamic split conjecture together, then quantum gravity should seek not to impose black hole formulas on cosmology, but rather to explain how non-equilibrium memory and teleodynamic bias emerge from microscopic gravitational degrees of freedom in an expanding universe. We emphasize that the present Letter does not claim that teleodynamics is the unique possible description of cosmological horizons, but rather that it provides a minimal non-equilibrium extension whose stationary limit reproduces all known thermodynamic results while yielding controlled deviations in evolving spacetimes. \\

Also, it should be noted that the “microstate counting" that appears in the teleodynamic derivation of cosmological horizon entropy \eqref{eq:STD} is fundamentally different from the quantum microstate counting underlying black hole entropy in string theory \citep{sv8Mayerson:2020acj,sv9Sen:1999mg,qe3Sen:2012kpz,qe4Banerjee:2010qc,qe5Banerjee:2011jp,qe6Sen:2012cj,qe7Dabholkar:2014ema}. Note that teleodynamics counts coarse-grained, time dependent correlation cells $N_{\rm TD}(t)=A_{\rm hor}(t)/\ell_\chi^2(t)$ on a dynamical FLRW horizon and incorporates non-equilibrium memory production through $S_{\rm TD}(t)=N_{\rm TD}(t)s_{\rm cell}(t)+\int^t\sigma_{\rm TD}(t')dt'$, whereas string theory counts genuine quantum states in a stationary Hilbert-space sector labeled by fixed charges $(M,J,Q)$ and protected by supersymmetry, modular invariance, and the presence of a timelike Killing vector. Because the universe possesses no such stationary structure—its $H(t)$ evolves, its correlation lengths grow, and $\sigma_{\rm TD}(t)>0$ reflects genuine gravitational memory. This also highlights that any future quantum gravity microstate construction capable of describing cosmology cannot reproduce the Bekenstein–Hawking formula as its semiclassical limit. Instead, it must reduce to the teleodynamic entropy. In this sense, the correct quantum gravity "microstates of the universe" might have to encode non-equilibrium horizon memory rather than stationary black hole degeneracy. This points to a profound shift in how quantum gravity should approach cosmological horizons and provides a deep, far-reaching interpretation of the thermodynamic split conjecture as well \citep{tsc1Trivedi:2025uup,tsc2Trivedi:2025ois}. \\

\textit{Acknowledgments}: The work of OT is supported in part by the Vanderbilt Discovery Doctoral Fellowship. The authors thank Robert Scherrer, Abraham Loeb, Sunny Vagnozzi, and Sergey Odintsov for their very helpful comments on Cosmological Teleodynamics. VV thanks Julia Velkovska and her colleagues in the Department of Physics and Astronomy at Vanderbilt University for their kind invitation to present the 2025 Wendell G. Holladay Lecture, which resulted in this paper.

	\bibliographystyle{cas-model2-names}
	\bibliography{cas-refs} 

\end{document}